\documentstyle[epsf,12pt]{article}
\hoffset = - 1 truecm
\def\beq{\begin{equation}}   \def\eeq{\end{equation}}
\def\bea{\begin{eqnarray}}  \def\eea{\end{eqnarray}} 
\def\noi{\noindent} \def\beeq{\begin{eqnarray}}
\def\eeeq{\end{eqnarray}}
\def\lsim{\raise0.3ex\hbox{$<$\kern-0.75em\raise-1.1ex\hbox{$\sim$}}}
\def\gsim{\raise0.3ex\hbox{$>$\kern-0.75em\raise-1.1ex\hbox{$\sim$}}}

\newcommand\mysection{\setcounter{equation}{0}\section}

\renewcommand{\theequation}{\thesection.\arabic{equation}}
\newcounter{hran} \renewcommand{\thehran}{\thesection.\arabic{hran}}

\def\bmini{\setcounter{hran}{\value{equation}}
\refstepcounter{hran}\setcounter{equation}{0}
\renewcommand{\theequation}{\thehran\alph{equation}}\begin{eqnarray}}

\def\bminiG#1{\setcounter{hran}{\value{equation}}
\refstepcounter{hran}\setcounter{equation}{-1}
\renewcommand{\theequation}{\thehran\alph{equation}}
\refstepcounter{equation}\label{#1}\begin{eqnarray}}

\def\emini{\end{eqnarray}\relax\setcounter{equation}{\value{hran}}\renewcommand{\theequation}{\thesection.\arabic{equation}}}

\def\ben{\begin{enumerate}}  \def\een{\end{enumerate}}

\def\cite#1{[\ref{#1}]} 
\def\citd#1#2{[\ref{#1}, \ref{#2}]}
\def\citt#1#2#3{[\ref{#1}, \ref{#2}, \ref{#3}]}
\def\citm#1#2{[\ref{#1}--\ref{#2}]}

\begin{document} 
\begin{center} 
\vbox to 1 truecm {} {\Large \bf Brane Universes and the Cosmological
Constant}

\vskip 1 truecm {\bf Ulrich Ellwanger} \vskip 3 truemm

{\it Laboratoire de Physique Th\'eorique\footnote{Unit\'e Mixte de
Recherche CNRS - UMR N$^{\circ}$ 8627},\\ Universit\'e Paris XI,
B\^atiment 210, 91405 Orsay Cedex, France\\ E-mail:
Ulrich.Ellwanger@th.u-psud.fr} \end{center} 
\vskip 1.5 truecm

\centerline{\bf  Abstract}
\vskip 5 truemm
The cosmological constant problem and brane universes are reviewed
briefly. We discuss how the cosmological constant problem manifests
itself in various scenarios for brane universes. We review attempts --
and their difficulties -- that aim at a solution of the cosmological
constant problem.

\vfill

\noindent LPT Orsay 05-51 \par
\noindent July 2005\par

\newpage \pagestyle{plain} \baselineskip=18 pt

\mysection{Introduction}

The smallness of the cosmological constant (or vacuum energy, or dark
energy) is one of the major puzzles of present day cosmology. The fact
that recent astronomical observations can be interpreted as evidence
for a nonvanishing cosmological constant does by no means
resolve the huge discrepancy between its measured value, and its value
computed in field theoretical models of particle physics. For recent
reviews of the subject see refs. \citm{1r}{8r}.

In the present paper we review different approaches to brane
universes, and discuss how the cosmological constant problem presents
itself. To this end we first have to clarify the nature of the problem
in the context of standard 4-dimensional cosmology (since later, in the
context of brane worlds, the observed acceleration rate of our universe
is no longer proportional to the vacuum energy located on the brane we
live on). We also give a short introduction to brane universes; here we
will be far from complete, but introduce just the formalisms required
subsequently. 

Then we review several brane world scenarious, that differ in the way
how the 4-dimen\-sional behaviour of gravity is ensured (at least over
the tested range of length scales). We will give the arguments in
favour of possible solutions of the cosmological constant problem in
some of these models, but later we have to conclude that none of these
attempts proves to be successful at present.

\mysection{The Cosmological Constant Problem}

The origin of the cosmological constant problem is the application of
Einstein's equations for the metric $g_{\mu\nu}(x)$ to the cosmological
evolution of our universe, that otherwise gives rise to the very
successful cosmological standard model. These equations involve the
Ricci tensor $R_{\mu\nu}$ and the Ricci scalar $R$ constructed from the
metric:
\beq \label{1.1e} R_{\mu\nu} - {1 \over 2} \ g_{\mu\nu} \ R = - \kappa
\ T_{\mu\nu} \eeq
\noi with $\kappa = 8 \pi {\cal G}/c^2 \cong 1,865 \cdot 10^{-29} \
{\rm m/g}$. $T_{\mu\nu}$ in (\ref{1.1e}) is the energy-momentum tensor
of matter which acts as source for the gravitational field. Well-known
vacuum solutions of eq. (\ref{1.1e}) (with $T_{\mu\nu}=0$) are the
Schwartzschild solution, that gives rise to the gravitational
attraction between massive objects, and gravitational waves, that await
their discovery.

In order to apply eq. (\ref{1.1e}) to the cosmological evolution of our
universe one can use the observational fact that, at cosmological
scales, the universe can be considered as homogenous and isotropic.
Then matter can be modelled by a (possibly relativistic) homogenous and
isotropic perfect fluid with matter density $\rho(t)$ (=$T_{00}(t)$)
and pressure $p(t)$ (=$T_{ii}(t)$, no sum over $i$). The equation of
state  of a perfect fluid determines a relation $p = p(\rho )$. Usually
one assumes $ p = w \rho $, where the constant $w$ depends on the
microscopic properties of the fluid: for nonrelativistic matter one has
$w \simeq 0$, whereas for relativistic matter (radiation) one has
$w \simeq \frac{1}{3}$.

For the metric $g_{\mu\nu}(x)$ on can make the Robertson-Walker ansatz
\beq \label{1.2e} ds^2 = dt^2 - a^2(t) \left ( {dr^2 \over 1 - kr^2} +
r^2 \left ( d \theta^2 + \sin^2 \theta d\phi^2 \right ) \right ) \ .
\eeq

\noi Here the constant $k$ can be chosen, after an appropriate
rescaling of $r$ and $a(t)$, as $k = 0,\ \pm 1$. It determines the
global geometry of three-dimensional space, which is flat for $k =
0$, a three-dimensional hypersphere (and hence closed) for $k = 1$, and
hyperbolic (open) for $k = - 1$.  $a(t)$ in (\ref{1.2e}) is the scale
factor of the three-dimensional space, and its time dependence is
determined by the 00 or $ii$ components of Einstein equations
(\ref{1.1e}):
\bminiG{1.3e}
\label{1.3ae}
3 \ {\dot{a}^2 + k \over a^2} = \kappa \ \rho(t) \ ,
\eeeq
\beeq
  \label{1.3be}
  - \ {2a \ddot{a} + \dot{a}^2 + k \over a^2} = \kappa \ p(t)
\emini
\noi where $\dot{a} = da/dt$. 
(Evidently eqs. (\ref{1.3e}) imply some
relation between $\rho$ and $p$ which implies, with $p=w\rho$,
$\dot{\rho} + 3 \ {\dot{a} \over a} (1 + w )\rho = 0$ corresponding to
energy conservation.)

In general, once the dynamics of matter (in the form of fields) 
is described by a Lagrangian ${\cal L}_M$ minimally coupled to the
metric $g_{\mu\nu}$ (in order to ensure invariance under general
coordinate transformations), the energy momentum tensor is
obtained as the variation of ${\cal L}_M$ with respect to $g_{\mu\nu}$:
\beq \label{1.4e} T_{\mu\nu}(x) = {2 \over \sqrt{-g(x)}} \ {\delta \over
\delta g^{\mu\nu}(x)} \int d^4 x' \ {\cal L}_M (x') \ . \eeq

In the case of homogenous and isotropic field configurations, the
general structure of $T_{\mu\nu}(x)$ is then
\beq \label{1.5e} T_{00} = \rho (t) + \Lambda \quad , \qquad T_{ii} =
p(t) - \Lambda \ .\eeq

Typical contributions to the (cosmological) constant $\Lambda$ arise
from the effective (classical or quantum) potential in ${\cal L}_M$,
\beq \label{1.6e} \Lambda = V_{eff}(\widehat{\phi})\ , \eeq
\noi where $\widehat{\phi}$ are the fields at the minimum of $V_{eff}$.
Now eqs. (\ref{1.3e}) are replaced by
\bminiG{1.7e}
\label{1.7ae}
3 \ {\dot{a}^2 + k \over a^2} = \kappa \ (\rho(t)+\Lambda) \ ,
\eeeq
\beeq \label{1.7be}
  - \ {2a \ddot{a} + \dot{a}^2 + k \over a^2} = \kappa \ (p(t)-\Lambda)
\ .\emini

\noi These equations have to be confronted with astronomical
measurements, notably i) the Hubble constant $H_0 =
\frac{\dot{a}}{a}|_{today}$ = redshift vs. distance, ii) the present
acceleration $\frac{\ddot{a}}{a}$ of the universe (from distant
supernovae counting), iii) the microwave background anisotropy and iv)
dynamical matter measurements. Details of the comparison of these
measurements with eqs. (\ref{1.7e}) are discussed in refs.
\citm{1r}{8r}, here we will just present the results:

First, all measurements are compatible with a flat universe ($k = 0$),
and the absence of a large component of radiation ($p(t) \simeq 0$ in
eq. (\ref{1.7be})). For the present Hubble constant, non-relativistic
matter density $\rho_0$ and the cosmological constant $\Lambda$ the
best fits give
\beq \label{1.8e}
H_0 \simeq 65\frac{km}{sec}(Mpc)^{-1}\ ,\quad
\rho_0 \simeq 3\cdot 10^{-24}\frac{g}{m^3}\ ,
\quad \Lambda \simeq 6\cdot 10^{-24}\frac{g}{m^3} 
\sim (2.3\cdot 10^{-3}eV)^4\ .
\eeq

\noi Whereas the value for $\rho_0$ indicates that there seems to be
more dark matter than visible matter (in agreement with other
observations), even the order of magnitude of $\Lambda$ is difficult to
understand: From the standard model of particle physics we expect
$\Lambda \sim (1\ GeV)^4$ from QCD, $\Lambda\sim (100\ GeV)^4$ from the
Higgs potential, and even $\Lambda \sim (10^{18}\ GeV)^4$ from quantum
contributions to the vacuum energy, with an ultraviolet cutoff of the
order of the Planck scale. This latter value can be improved in the
case of supersymmetric extensions of the standard model of particle
physics: If supersymmetry is broken at a scale $M_{SUSY}\ \sim\ 100\
GeV$, quantum corrections to the vacuum energy give $\Lambda\sim (100\
GeV)^2 \cdot(10^{18}\ GeV)^2$ or $\Lambda\sim (100\ GeV)^4$, depending
on the way supersymmetry breaking manifests itself in the particle
spectrum (most models would lead to the first larger value; for a
discussion and alternatives see, e.g., ref. \cite{uecc}). In any
case, the discrepancy between these values and the one given in eq.
(\ref{1.8e}) is enormous, which is what one denotes as the cosmological
constant problem.

An important point is to be made here, however: the "observed" value of
$\Lambda$ is deduced from eqs. (\ref{1.7e}), essentially from the
measurement of the present acceleration $\frac{\ddot{a}}{a}$ of the
universe. Hence, possible solutions to this puzzle could be related to
the fact that eqs. (\ref{1.7e}) are too naive. In fact, these equations
get modified in higher dimensional universes, notably in brane
universes, which are the subject of the next chapter.

\mysection{Brane Universes}

The Einstein equations (\ref{1.1e}) can trivially be generalized to
space-times with $D > 4$ dimensions. Nowadays there exist two concepts
that can nevertheless lead to an effective 4-dimensional behaviour of
gravity, at least over the range of length scales (from millimeters to
the actual size of the universe) where 4-d gravity agrees with
experiment. The traditional concept is the one of Kaluza and Klein,
according to which all extra $D-4$ dimensions are compact. It ensures
also a 4-d behaviour of all other fundamental interactions over the
range they are tested, provided the size of the extra dimensions is
less than $\sim (100\ GeV)^{-1}$ (in units where $c=\hbar=1$).

The new concept is the one of brane universes, which are motivated by
the presence of D-branes in string theory. Here matter -- at least the
observed fields of the standard model of particle physics -- live on a
$3+1$ dimensional hypersurface that is embedded in a larger
D-dimensional space-time, traditionally denoted as bulk. (A p-brane
is a brane with p spatial dimension, thus our $3+1$ dimensional
space-time corresponds to a 3-brane.) Now a 4-d behaviour of the other
fundamental interactions is guaranteed, but a 4-d behaviour of gravity
-- at least over the required range of scales -- is not automatic. In
the subsequent sections of this paper we will review different brane
world models that do lead to 4-d gravity, and discuss the cosmological
constant problem in each of them.

First, however, we will present briefly how brane worlds are described
in terms of a gravitational action and gravitational field equations.

In $D=4+N$ space-time dimensions we will split the coordinates into
$x^\mu$, $\mu = 1\dots 4$, and $y^\alpha$, $\alpha = 1\dots N$. The
indices of the metric tensor are split correspondingly, hence
$g^{(4+N)}$ has indices $g_{\mu\nu} = g_{\mu\nu}^{(4)}$, $g_{\mu\alpha}$
and $g_{\alpha\beta}$.

For our subsequent purposes it is convenient to assume that the brane is
located at $y^\alpha = 0$. (This assumption is not invariant under
general coordinate transformations, but corresponds to certain "gauge".
This fact has to be considered carefully once one wants to identify the
full set of physical fluctuations of the metric. Also, in the case of
inhomogenously distributed matter on the brane, this is not necessarily
the most convenient choice of gauge \cite{gt}.)

In the action below we will assume that the bulk is empty, up to a
(cosmological) constant $\Lambda_{Bulk}$. On the brane we will allow
for a general Lagrangian ${\cal L}_M$, that can include a (different) 
cosmological constant $\Lambda$ as well. Then the action reads
\beq \label{2.1e}
S = \int d^4 x d^N y
\left[\sqrt{-g^{(4+N)}}\left(\frac{1}{2\kappa_B} R^{(4+N)} +
\Lambda_{Bulk}\right)
 + \sqrt{-g^{(4)}} \delta^N (y^{\alpha} ) {\cal L}_{M}\right]
 \eeq
where $\kappa_B$ is the "fundamental" gravitational coupling constant in
the bulk.

The effect of the "brane" term $\sim \delta^N (y^{\alpha} )$ on the
gravitational field equations can most easily be studied by replacing
${\cal L}_{M}$ by $\Lambda$, and assuming one extra dimension $y$ only.
Then the components $\mu\nu$  of the Einstein equations (\ref{1.1e})
become
\beq \label{2.2e}
R^{(4+1)}_{\mu\nu} - \frac{1}{2} g_{\mu\nu} R^{(4+1)} 
= \kappa_B \left(g_{\mu\nu} \Lambda_{Bulk} + \frac{1}{\sqrt{g_{yy}}}
\delta(y) g_{\mu\nu} \Lambda\ \right) . \eeq

In order to match the Dirac $\delta$-function on the right hand side of
eq. (\ref{2.2e}) it is useful to recall that its left hand side, to
linear order in $g_{\mu\nu} - \eta_{\mu\nu}$, reads $\Box^{(4)}
g_{\mu\nu} +  \partial_y \partial^y g_{\mu\nu}$. The second derivatives
w.r.t. $y$ generate a corresponding $\delta$-function provided that
\beq \label{2.3e}
\partial_y g_{\mu\nu} (y = +\epsilon) = \partial_y g_{\mu\nu} (y =
- \epsilon) + \frac{\kappa_B \Lambda}{2 \sqrt{g_{yy}}} g_{\mu\nu}\ .
\eeq

Hence, the first derivative of $g_{\mu\nu}$ w.r.t. $y$ has to jump
across the brane, by an amount depending on the cosmological constant
$\Lambda$ on the brane (which corresponds to what is also called the
brane tension).

How does the cosmological constant problem represent itself in brane
universes? First we have to recall that the values for both  $H_0$ and
$\Lambda$ in (\ref{1.8e}), deduced from eqs. (\ref{1.7e}), are tiny
compared to "fundamental" scales $(1\ GeV)^{-1}$ or $(100\ GeV)^{-1}$
in particle physics. The reason therefore are the relatively tiny
measured values for the expansion rate $\frac{\dot a}{a}$ and the
acceleration $\frac{\ddot a}{a}$ of the universe. Hence, as compared to
fundamental scales in particle physics, our universe can be considered
as practically static (time independent). Hence realistic brane
universes, apart from leading to an effective 4-d behaviour of gravity,
should allow for a time independent scale factor on the brane in zeroth
approximation. Only subsequently one has to check, whether a relatively
small amount of additional matter and vacuum energy on the brane
generates a relatively slow cosmological evolution in agreement with
observations.

Now recall that both classical and quantum fields confined to a brane
tend to generate a "large" cosmological constant $\Lambda$ {\it on the
brane}. Is it possible that static brane worlds exist for an
{\it arbitrary} cosmological constant $\Lambda$ on the brane? This
possibility is not excluded, given lower dimensional examples: The
Schwartzschild solution in $d=4$ can be interpreted as a 0-brane in a
4-dimensional bulk, and it is know to be both static and stable for
arbitrary mass parameter m that plays the role of $\Lambda$ on the
brane. Also, the cosmic string solution in $d=4$ can be considered as a
1-brane embedded in a higher dimensional bulk, and is static for
arbitrary string tension (but possibly unstable \cite{gl}). Naive
extrapolations of these configurations to $4+N$ dimensional brane worlds
have, however, to be modified in order to generate an effective 4-d
behaviour of gravity. As we will discuss below, these realistic
scenarios do no longer seem to allow for an arbitrary value of
$\Lambda$.

\mysection{Effective 4-d Gravity versus the Effective Cosmological
Constant}

Effective 4-d gravity corresponds to an (asymptotic) gravitational 
potential $V(r) \sim \frac{1}{r}$, and {\it not}  $V(r) \sim
\frac{1}{r^{1+N}}$ in a $D=4+N$ dimensional brane world. The desired
gravitational potential is generated by the exchange of a massless
graviton, that satisfies the {\it four dimensional} wave equation. Let
us now consider various brane world scenarios that lead to such a
gravitational field equation.

\subsection {One compact extra dimension ($D=5$)} 

Here the only extra dimension $y$ is considered as periodic, i.e. the
points $y$ and $y+2\pi R$ are identified and the metric (and its
derivatives) have to satisfy corresponding periodicity conditions in
$y$. The search for corresponding cosmological solutions of Einstein's
equations is greatly simplified by the fact that due to the assumptions
of homogeneity {\it and} time independence the metric $g_{\mu\nu}$
depends on $y$ only. Hence Einstein's equations for $g_{\mu\nu}$ 
become a simple second order differential equation in $y$ where,
however, both the boundary condition (\ref{2.3e}) and the periodicity
condition have to be imposed. One finds rapidly that these conditions
are in conflict \cite{bdl} unless a) a second brane with tension
$\Lambda_2 = -\Lambda$ is introduced, and b) the brane cosmological
constant $\Lambda$, the bulk cosmological constant $\Lambda_{Bulk}$ and
the 5-d gravitational constant $\kappa_B$ are related via
\beq \label{3.1e}
\Lambda_{Bulk} = -\frac{1}{6}\kappa_B^2\Lambda^2\ .
\eeq
For nonvanishing $\Lambda_{Bulk}$ and $\Lambda$ satisfying (\ref{3.1e})
this brane world is named Randall-Sundrum scenario I \cite{rs1}.

The way 4-d gravity emerges in this scenario is fairly straightforeward
due to the compactness of the extra dimension: First, one has to
decompose the metric into a background metric $g_{\mu\nu}^0(y)$ and
fluctuations $h_{\mu\nu}$:
\beq \label{3.2e}
g_{\mu\nu}(x,y) = g_{\mu\nu}^0(y) + h_{\mu\nu}(x,y)
\eeq

Then, the linearized Einstein's equations (in $h_{\mu\nu}$) become a
wave equation for $h_{\mu\nu}$:
\beq \label{3.3e}
\Box^{(4)}h_{\mu\nu}+ O^{(y) \rho
\phantom{xx} \sigma}_{\phantom{(o) \rho} \mu\nu} h_{\rho \sigma} = 0
\eeq
with
\beq \label{3.4e}
O^{(y) \rho
\phantom{xx} \sigma}_{\phantom{(o) \rho} \mu\nu} h_{\rho \sigma} =
\partial_y \partial^y h_{\mu\nu} - 2 R^{(0) \rho
\phantom{xx} \sigma}_{\phantom{(o) \rho} \mu\nu} h_{\rho \sigma}\ ,
\eeq
where $R^{(0) \rho \phantom{xx} \sigma}_{\phantom{(o) \rho} \mu\nu}$ is
the Riemann tensor constructed from the background metric 
$g_{\mu\nu}^0(y)$.

For a compact extra dimension the spectrum of $O^{(y)}$ is
semi-positive and discrete: \break
$O^{(y)} h^{(n)}_{\mu\nu} =
k_n^2 h^{(n)}_{\mu\nu}$ with $k_0^2 = 0$, $k_n^2 > 0$ for $n \geq 1$.
The zero mode  $h_{\mu\nu}^0$ represents the massless 4-d graviton
(obeying the wave equation $\Box^{(4)} h_{\mu\nu}^{0} = 0$), and its
exchange generates the desired 4-d gravitational potential
$V(r)\sim\frac{1}{r}$.

On the other hand, the compactness of the extra dimension was also
responsable for the two constraints $\Lambda_2 = -\Lambda$ and eq.
(\ref{3.1e}) above. Recall that our present assumption of a time
independent background metric $g_{\mu\nu}^0$ corresponds to a vanishing
effective 4-d cosmological constant (eqs. (\ref{1.7e}) with $k=0$ and
vanishing right hand sides). Now this assumption does not require a
vanishing cosmological constant on the brane(s), but instead two
fine tuning conditions on the fundamental parameters of the model. In
fact, if both of these conditions are violated, one obtains not only a
rapidly accelerating scale factor on the brane(s), but also a time
dependent effective 4-d gravitational constant in strong disagreement
with observations. Hence the cosmological constant problem is far from
being solved, in some sense the need to satisfy two fine tuning
conditions among the fundamental parameters is even worse than before.
As discussed in \cite{const} this remains valid after adding scalar
fields with arbitrary Lagrangian in the bulk.

\subsection{One Non-Compact Extra Dimension ($D=5$)}

In \cite{rs2} it has been proposed to let the size of the extra
dimension go to infinity, whereupon it becomes non-compact. The
periodicity condition (in $y$) on the metric is now replaced by the
condition that $g_{\mu\nu}(y)$ remains finite for $y \to \pm \infty$.
Amazingly, 4-d gravity still emerges down to sufficiently small length
scales:

Now, the spectrum of the operator $O^{(y)}$ in (\ref{3.3e}) and
(\ref{3.4e}) is continuous, but the corresponding eigenfunctions are
not normalizable with the exception of one normalizable zero mode. As
shown in \cite{rs2}, the existence of this normalizable zero mode is
sufficient to generate 4-d gravity down to sufficiently small length
scales (with deviations at small distances that are possibly
measurable in the future).

Does the absence of periodicity conditions on the metric lead to the
absence of the two fine tuning conditions before? Unfortunately not,
since the condition that $g_{\mu\nu}(y)$ remains finite for $y \to \pm
\infty$ still requires a particular relation among the fundamental
parameters, that turns out to be identical to eq. (\ref{3.1e}). Hence
the number of fine tuning conditions is reduced from 2 to 1 (due to the
absence of the second brane), but the remaining condition (\ref{3.1e})
is still required in order to tune the effective 4-d cosmological
constant to zero.

\subsection{Several Non-Compact Extra Dimensions (the DGP
Model)}

A scenario for effective 4-d gravity with non-compact extra
dimensions, but no cosmological constant in the bulk, has been proposed
in \citt{dgp}{cohol}{dg}. Instead, an additional Einstein term (that
includes a kinetic term for the graviton) is added to the action on the
brane. Replacing ${\cal L}_M$ in (\ref{2.1e}) by $\Lambda$ and omitting
the cosmological constant in the bulk, the action is given by
\beq \label{3.5e}
S=\int d^4 x d^N y \left[ \sqrt{-g^{(4+N)}} \frac{1}{2 \kappa_B}
R^{(4+N)} 
+ \sqrt{-g^{(4)}} \delta^N (y) \left( \frac{1}{2 \kappa_4}
R^{(4)} + \Lambda \right) \right]
\eeq
which involves two different gravitational couplings $\kappa_B$ (in
the bulk) and $\kappa_4$ (on the 3+1 dim. brane). It can even be
argued  that it would be unnatural to omit the Einstein term $R^{(4)}$
on the brane, since matter induced quantum corrections would generate
it anyhow. In \citd{dgp}{cohol} the above model has been formulated in
$D=5$ dimensions ($N=1$), but an interesting result with respect to
the cosmological constant problem emerges only in $D \geq 6$
dimensions ($N \geq 2$) as considered in \cite{dg}.

The action (\ref{3.5e}) leads to an interesting structure for the
graviton propagator $G(p, |y-y'|)$, where $p$ is the 4-d momentum
(parallel to the brane), and $y$, $y'$ are arbitrary end points in the
bulk. (Here we neglect for simplicity the tensorial structure of the
graviton propagator.) Writing $\kappa_4^{-1} = M_{Pl}^2$, 
$\kappa_B^{-1} = M_B^{N+2}$ (in $D=4+N$ dimensions), one finds for $N
> 2$
\beq \label{3.6e}
G(p,|y-y'|) \simeq \frac{|y-y'|^{2-N} M_B^{N-2}}{M_B^2 + p^2 M_{Pl}^2
|y-y'|^{2-N} M_B^N}\ .
\eeq

We recall that 4-d gravity requires a "brane-to-brane" propagator
$G(p,0)$ that behaves like $\frac{1}{p^2}$. The limit $|y-y'| \to 0$
in eq. (\ref{3.6e}) is obviously singular for $N > 2$. 
(A detailed study for $N=2$ reveals the appearance of
logarithmic singularities in this case.)
A naive
regularization consists in replacing $|y-y'|^{2-N}$, for $|y-y'| \to
0$, by $M_B^{N-2}$, the (fundamental) gravitational constant in the
bulk. Then one obtains
\beq \label{3.7e}
G(p,0) \simeq \frac{1}{M_B^2 + p^2 M_{Pl}^2 / M_B^2}
\eeq
which behaves like $\frac{1}{p^2}$ only for $p^2 \gg M_B^4/M_{Pl}^2$, 
i.e. for length scales $r < r_c \sim \frac{M_{Pl}}{M_B^2}$. In order not
to mess up the successful predictions of the cosmological standard
model, we need $r_c\ \gsim\ H_0^{-1} \sim 10^{33}\ eV^{-1}$ which
corresponds to $M_B\ \lsim\ 10^{-3}\ eV$, indeed a quite unusual value
for the "fundamental" (D-dimensional) gravitational scale.

The essential feature of the model is then the modified (massive)
behaviour of the graviton propagator at length scales larger than
$r_c$. As a consequence, gravity does not necessarily react to sources
that are smooth at scales larger than $r_c$ $\sim\ H_0^{-1}$, as it is
the case for a cosmological constant $\Lambda$ on the brane. As
proposed in \cite{dg} and \cite{dgs} (see ref. \cite{g} for a review),
this could lead to a potential solution of the cosmological constant
problem.

However, the scenario faces two severe problems: First, less naive UV
regularizations of the graviton propagator (by smearing out the
previously vanishing width of the brane) tend to be in conflict with
the requirement to solve Einstein's equations also at small distances.
This can be shown explicitely in a solvable scenario where the
previously infinitely thin brane is considered as "hollow", and
Einstein's equations are required to be satisfied also inside as well
as across the surface \cite{ue}. As a consequence, additional
constraints on the parameters of the model appear, that re-introduce
a fine tuning condition for the existence of a static solution.

Second, a study of all tensorial components of the graviton propagator
reveals the presence of negative norm states (with negative residues)
\cite{nns} which signal an instability of the configuration. Although
the details depend on the UV regularization mentionned above, the fact
that the effective mass term for the graviton is not of the Fierz-Pauli
form renders this problem certainly difficult to solve.
Hence it cannot be claimed at present that these scenarios solve the
cosmological constant problem.

\subsection{Two Compact Extra Dimensions}

Scenarios with two "football shaped" compact extra dimension and a time
independent metric for an arbitrary value of the cosmological constant
on the brane have been proposed in \citd{footb1}{footb2}. Due to the
compactness of the extra dimensions the emergence of 4-d gravity is
evident, but the issue is now whether the junction conditions of the
metric on the brane (conditions on its derivatives perpendicular to the
brane, in analogy to the condition (\ref{2.3e}) in the case of one
co-dimension $N=1$) can be satisfied for $\Lambda$ arbitrary. 

This requires a particular form of the curvature in the bulk, which is
induced by additional matter in the form of a $U(1)$ gauge field with
field strength $F_{MN}$ in the bulk. A configuration of
$F_{MN}$ (with indices in the extra dimensions) that solves the gauge
field equations of motion, generates the required curvature and is
adopted to the singular geometry at the position of the brane
corresponds to the one of a magnetic monopole, leading to an arbitrary
value of the deficit angle $\alpha$ (w.r.t. $2\pi$) of a circle that
encloses the brane in the two extra dimensions. A priori $\alpha$ can
be chosen such that the junction conditions are satisfied for any value
of $\Lambda$, and then the scale factor on the brane is time
independent for any value of $\Lambda$, which seems to solve the
cosmological constant problem.

However, as noted in \citd{footb2}{footb3}, the magnetic flux
corresponding to $F_{MN}$ has to satisfy a Dirac quantization
condition such that it is an integer devided by two times the gauge
coupling. It follows that only discrete values of the deficit angle
$\alpha$, and hence of $\Lambda$ are allowed. This -- and the
instability with respect to perturbations \cite{vc1} -- rules out the
possibility to solve the cosmological constant problem for arbitrary
(notably time dependent, as near the end of an inflationary epoch)
values of $\Lambda$.

\subsection{Self Tuning Models}

In fact, this previous scenario is just a particular (though a priori
quite promising) case of so-called self tuning models, that are
reviewed in \cite{st}. Their common feature is the presence of
additional fields in the bulk, whose "vacuum" configurations can be
arranged such that all equations of motion are satisfied, and the
metric and all fields (and hence the scale factor on the brane) are
time independent, for arbitrary values of $\Lambda$. 

The first models of this kind (with $N=1$ extra dimension) involved a
dilaton like scalar field, present also in the bulk, with an
exponential potential on the brane \cite{stm1}.
However, in the case of non compact extra dimensions these
models lead generically to metrics that are singular at infinity. As
shown in \cite{stm2}, regularization of these singularities
re-introduces fine tuning.

In the case of compact extra dimensions, flux quantization conditions
(as the one above) allow generically just for discrete sets of
parameters. This does not allow for reasonable 4-d physics, as
clarified in \cite{st}. In addition, particular (fine tuned)
cosmological initial conditions are required in these scenarios, which
corresponds just to a shift of the cosmological constant problem and
not to its solution.

\mysection{Outlook}
It is certainly true that brane worlds offer new tools that could
potentially solve the cosmological constant problem. However, from a
4-d point of view, it seems that a solution of the problem is
practically impossible \cite{1r} as long as neither gravity itself,
nor the way gravity reacts to energy-momentum, is modified. Note that
modifications at small distances (as expected anyhow, if
gravitational quantum UV divergencies are regularized) are useless here,
since the problem -- the very slow evolution of the universe as compared
to fundamental scales in particle physics -- concerns physics at very
large distances (or time scales). 

On the other hand, any brane world scenario that is proposed as a
solution of the problem must also lead to an effective 4-d theory of
gravity and its coupling to matter (which is, after all, what we see).
It seems that such an effective 4-d theory must have unconventional
properties at (very) large distances. It is notoriously difficult to
tamper with the infrared behaviour of gauge theories such as general
relativity without generating inconsistencies as negative norm states,
tachyons, or violations of unitarity and/or causality. Nevertheless it
is not excluded that, by looking for solutions of the cosmological
constant problem, further studies of brane worlds uncover consistent
modifications of 4-d gravity at large distances that lead to a solution
of the perhaps most puzzling problem of fundamental physics today.

\newpage 
%\vskip 2 cm

\def\labelenumi{[\arabic{enumi}]} 
\noindent {\large\bf References} 
\ben

\item\label{1r} S. Weinberg, Rev. Mod. Phys. {\bf 61} (1989) 1

\item\label{2r} V.~Sahni and A.~A.~Starobinsky,
  Int.\ J.\ Mod.\ Phys.\ D {\bf 9} (2000) 373
  [arXiv:astro-ph/9904398].

\item\label{3r}  E.~Witten,
``The cosmological constant from the viewpoint of string theory,'' 
arXiv:hep-ph/0002297.

\item\label{4r} S.~M.~Carroll,
  Living Rev.\ Rel.\  {\bf 4} (2001) 1
  [arXiv:astro-ph/0004075].

\item\label{5r} S.~Weinberg,
``The cosmological constant problems,''
 arXiv:astro-ph/0005265.

\item\label{6r} U.~Ellwanger, 
  ``The cosmological constant,''
arXiv:hep-ph/0203252.

\item\label{7r} P.~J.~E.~Peebles and B.~Ratra,
  Rev.\ Mod.\ Phys.\  {\bf 75} (2003) 559
  [arXiv:astro-ph/0207347].
  
\item\label{pad} T.~Padmanabhan,
  Phys.\ Rept.\  {\bf 380} (2003) 235
  [arXiv:hep-th/0212290].

\item\label{8r} S.~M.~Carroll, eConf {\bf C0307282} (2003) TTH09
  [AIP Conf.\ Proc.\  {\bf 743} (2005) 16]
  [arXiv:astro-ph/0310342].
  
\item\label{uecc} U.~Ellwanger,
  Phys.\ Lett.\ B {\bf 349} (1995) 57
  [arXiv:hep-ph/9501227].
  
\item\label{gt}  J.~Garriga and T.~Tanaka,
  Phys.\ Rev.\ Lett.\  {\bf 84} (2000) 2778
  [arXiv:hep-th/9911055]. 

\item\label{gl}  R.~Gregory and R.~Laflamme,
  Phys.\ Rev.\ Lett.\  {\bf 70} (1993) 2837
  [arXiv:hep-th/9301052].
  
\item\label{bdl}  P.~Binetruy, C.~Deffayet and D.~Langlois,
  Nucl.\ Phys.\ B {\bf 565} (2000) 269
  [arXiv:hep-th/9905012].
  
\item\label{rs1} L.~Randall and R.~Sundrum,
  Phys.\ Rev.\ Lett.\  {\bf 83} (1999) 3370
  [arXiv:hep-ph/9905221].
  
\item\label{const} U.~Ellwanger,
  Phys.\ Lett.\ B {\bf 473} (2000) 233
  [arXiv:hep-th/9909103].
  
\item\label{rs2}   L.~Randall and R.~Sundrum,
  Phys.\ Rev.\ Lett.\  {\bf 83} (1999) 4690
  [arXiv:hep-th/9906064].

\item\label{dgp} G.~R.~Dvali, G.~Gabadadze and M.~Porrati,
  Phys.\ Lett.\ B {\bf 485} (2000) 208
  [arXiv:hep-th/0005016].

\item\label{cohol} H.~Collins and B.~Holdom,
  Phys.\ Rev.\ D {\bf 62} (2000) 124008
  [arXiv:hep-th/0006158].

\item\label{dg}  G.~R.~Dvali and G.~Gabadadze,
  Phys.\ Rev.\ D {\bf 63} (2001) 065007
  [arXiv:hep-th/0008054].
  
\item\label{dgs} G.~Dvali, G.~Gabadadze and M.~Shifman,
  Phys.\ Rev.\ D {\bf 67} (2003) 044020
  [arXiv:hep-th/0202174].
  
\item\label{g} G.~Gabadadze,
  ``Looking at the cosmological constant from infinite-volume bulk,''
  arXiv:hep-th/0408118.
  
\item\label{ue}  U.~Ellwanger,
  JCAP {\bf 0311} (2003) 013
  [arXiv:hep-th/0304057].

\item\label{nns} 
 C.~Middleton and G.~Siopsis,
  ``Fat branes in infinite volume extra space,''
  arXiv:hep-th/0210033,\\
  S.~L.~Dubovsky and V.~A.~Rubakov,
  Phys.\ Rev.\ D {\bf 67} (2003) 104014
  [arXiv:hep-th/0212222].

\item\label{footb1}
  S.~M.~Carroll and M.~M.~Guica,
  arXiv:hep-th/0302067;\\
   I.~Navarro,
  JCAP {\bf 0309} (2003) 004
  [arXiv:hep-th/0302129].
  
\item\label{footb2} 
  Y.~Aghababaie {\it et al.},
  JHEP {\bf 0309} (2003) 037
  [arXiv:hep-th/0308064].

\item\label{footb3} 
   I.~Navarro,
  Class.\ Quant.\ Grav.\  {\bf 20} (2003) 3603
  [arXiv:hep-th/0305014].
  
\item\label{vc1} J.~Vinet and J.~M.~Cline,
  Phys.\ Rev.\ D {\bf 70} (2004) 083514
  [arXiv:hep-th/0406141],\\
  J.~Vinet and J.~M.~Cline,
  Phys.\ Rev.\ D {\bf 71} (2005) 064011
  [arXiv:hep-th/0501098].
  
\item\label{st}
   H.~P.~Nilles, A.~Papazoglou and G.~Tasinato,
  Nucl.\ Phys.\ B {\bf 677} (2004) 405
  [arXiv:hep-th/0309042].
  
\item\label{stm1}  N.~Arkani-Hamed, S.~Dimopoulos, N.~Kaloper and
R.~Sundrum,
  Phys.\ Lett.\ B {\bf 480} (2000) 193
  [arXiv:hep-th/0001197];\\
S.~Kachru, M.~B.~Schulz and E.~Silverstein,
  Phys.\ Rev.\ D {\bf 62} (2000) 045021
  [arXiv:hep-th/0001206].
  
\item\label{stm2}   S.~Forste, Z.~Lalak, S.~Lavignac and H.~P.~Nilles,
  Phys.\ Lett.\ B {\bf 481} (2000) 360
  [arXiv:hep-th/0002164];\\
    S.~Forste, Z.~Lalak, S.~Lavignac and H.~P.~Nilles,
  JHEP {\bf 0009} (2000) 034
  [arXiv:hep-th/0006139].

\een

\end{document}